\documentclass[slac_one]{revtex4}
\usepackage{graphicx}
\usepackage{epsfig}
\usepackage{epstopdf}
\usepackage{fancyhdr}
\begin{document}

\title{New MiniBooNE Results} 

%

\author{Zelimir Djurcic (for the MiniBooNE Collaboration)}
\affiliation{Department of Physics, Columbia University, New York, NY 10027, USA}

\begin{abstract}
The MiniBooNE experiment at Fermilab was designed to be a definitive test
of the LSND evidence for neutrino oscillations and has recently reported
first results of a search for electron-neutrino appearance in a
muon-neutrino Booster beam. No significant excess of events was observed at
higher energies, but a sizable excess of events was observed at lower
energies. The lack of the excess at higher energies allowed MiniBooNE to
rule out simple two-neutrino oscillations as an explanation of the LSND
signal. However, the excess at lower energies is presently unexplained. 

A new data set of neutrinos from the NuMI beam line measured with the
MiniBooNE detector at Fermilab has been analyzed.  The
measurement of NuMI neutrino interactions in MiniBooNE provide a clear
proof-of-principle of the off-axis beam concept that is planned to be used
by future neutrino experiments such as T2K and NOvA.
Moreover, it complements the
first oscillation results and will help to determine whether the
lower-energy excess is due to background or to new physics.

New results from the re-analysis of low energy excess from the Booster
beam line and the results from measurements of neutrino interactions from
NuMI beam line are discussed.
MiniBooNE observes an unexplained excess of $128.8 \pm 20.4 \pm 38.3$ electron-like events in the energy region
$200 < E_{\nu} < 475$ MeV. The NuMI data sample currently has a large systematic errors associated with  $\nu_{e}$ events,
but shows an indication of an excess.
\end{abstract}

\maketitle

\thispagestyle{fancy}

\section{Introduction}

The existence of neutrino oscillations have been confirmed in the results of
 solar-neutrino~\cite{solarE}, reactor-neutrino~\cite{reactorE}, 
atmospheric-neutrino~\cite{atmE}, and accelerator-neutrino~\cite{accE} experiments. 
These results implied the existence of two independent $\Delta m^2$ regions,
with  $\Delta m^2 \sim 8 \times 10^{-5}$ eV$^2$ in the solar, and
with $\Delta m^2 \sim 3 \times 10^{-3}$ eV$^2$ in the atmospheric sector.
The discovery of nonzero neutrino masses through the neutrino oscillations
has raised a number of very interesting questions about neutrinos and 
their connections to other areas of physics and astrophysics.
Unconfirmed evidence for neutrino oscillations, however, came from the LSND~\cite{lsndE}
experiment with $\Delta m^2$ at $\sim 1$ eV$^2$ value.
One question is whether there are sterile neutrinos at $\Delta m^2$ at $\sim 1$ eV$^2$ mass scale that do not 
participate in the standard weak interactions. 
This question is primarily being addressed by the MiniBooNE 
experiment.
The MiniBooNE experiment was designed to confirm or refute the LSND result
with higher statistics and different sources of systematic error.
If the LSND neutrino oscillation evidence was confirmed, it would, together
with solar, reactor, atmospheric and accelerator oscillation
data, imply Physics Beyond the Standard Model such as the existence
of light sterile neutrino~\cite{sorel}.
LSND observed an excess of $\bar\nu_{e}$ events in a 
$\bar\nu_{\mu}$ beam. MiniBooNE is performing both $\nu_{\mu} \rightarrow \nu_{e}$
and  $\bar{\nu}_{\mu} \rightarrow \bar{\nu}_{e}$ searches.
An additional data sample measured by the MiniBooNE detector comes from neutrinos
produced in the NuMI (Neutrinos from Main Injector) beam line.

\section{The MiniBooNE Experiment}

MiniBooNE is a fixed target experiment currently taking data at the Fermi National 
Accelerator Laboratory. The neutrino beam is produced from 8.89 GeV/c protons, from Fermilab Booster,
impinging on a 71 cm long and 1 cm diameter beryllium target. The target is located 
inside a magnetic focusing horn that increases the neutrino flux at the detector 
by a factor of $\sim$5, and can operate in both negative and positive polarities 
for $\nu$ and $\bar{\nu}$ running.  
MiniBooNE collected approximately $6.6 \times 10^{20}$ protons on target (POT)
in neutrino mode, and approximately $3.3 \times 10^{20}$ POT
in anti-neutrino mode, using the Booster neutrino beam (BNB).  These data samples are currently used in
the neutrino oscillation analysis. Only the neutrino analysis will be discussed in this report.
Results of a detailed anti-neutrino analysis are anticipated soon.

Mesons produced in the target decay-in-flight in a 50 m long decay pipe.
The neutrino beam is composed of $\nu_{\mu}$s from
$K^{+}/\pi^{+} \rightarrow \mu^{+} + \nu_{\mu}$
decays.  The neutrino beam propagates through  450 m of dirt before entering the detector.
There is a small contamination from $\nu_{e}$;
the processes that contribute to the intrinsic $\nu_{e}$
in the beam are $\mu^{+} \rightarrow e^{+} \nu_{e} \bar{\nu}_{\mu}$,
$K^{+} \rightarrow \pi^{0} e^{+} \nu_{e}$, 
and $K_{L}^{0} \rightarrow \pi^{\pm} e^{\pm} \nu_{e}$.
The flux modeling uses a {\sc Geant4}-based simulation of beam line geometry.
Hadron production in the target is based on the Sanford-Wang parametrization
of $p-Be$ cross-section, with parameters determined by a global fit
to $p-Be$ particle production data. The details are described in Ref.~\cite{MB_flux}.
Simulated neutrino flux has
an energy distribution with a peak at $E_{\nu} \sim 0.7$ GeV.
Therefore, the average L/E$_{\nu}$ ratio is $\sim$ 0.8 km/GeV
compared to LSND's L/E$_{\nu} \sim$ 1 km/GeV, where L
is the neutrino travel distance.

The neutrinos produced by the Booster are detected in the MiniBooNE 
detector~\cite{MB_detector} which is a 12.2~m 
spherical tank filled with 800 tons of pure mineral oil. 
The main MiniBooNE trigger is an accelerator signal indicating
a beam spill. Every beam trigger opens a 19.2 $\mathrm{\mu s}$ window
in which all events are recorded.
The time and charge of photomultiplier tubes (PMT) in the detector are used 
to reconstruct the interaction point, event time, energies, and particle tracks resulting from neutrino interactions. 
Neutrino interactions in the detector are simulated with the {\sc NUANCE} event generator package, 
with modifications to the quasi-elastic (QE) cross-section as described in~\cite{MB_ccqe}. Particles generated 
by {\sc NUANCE} are propagated through the detector, using a {\sc GEANT3}-based simulation which describes 
the emission of optical and near-UV photons via Cherenkov radiation and scintillation. Neutrino induced events are 
identified by requiring the event to occur during the beam spill, after rejection of cosmic ray muons and 
muon decay electrons. 

\section{First MiniBooNE Oscillation Result}

The MiniBooNE collaboration reported initial results of $\nu_{\mu} \rightarrow \nu_{e}$ search~\cite{MB_osc}
with the data sample of $5.58 \times 10^{20}$ POT.
Neutrino interactions are identified with the likelihood-based algorithm, where the event
parameters are varied to maximize the likelihood of the observed PMT hits.
 MiniBooNE conducted a blind analysis in order to complete an unbiased 
oscillation search. That means that the region where the oscillation
$\nu_{e}$ candidates were expected was closed for the analysis
until the reconstruction software and events selection cuts were finalized. 
After the analysis cuts were set, an oscillation analysis was performed in
the range of reconstructed neutrino energy, $475 < E_\nu < 1250$ MeV.
The estimated number of background events in the range, $475 < E_\nu < 1250$ MeV,
after the selection cuts were applied was $358\pm35(syst)$.
The estimate includes systematic uncertainties associated with neutrino flux,
neutrino cross sections, and the detector model.
The flux prediction has the uncertainties corresponding to the production
of $\pi$, $K$, and $K_L$ particles in the MiniBooNE target.
These uncertainties are quantified by a fit to external data sets
from previous experiments on meson production.
The cross section uncertainties are evaluated by
varying underlying cross section model parameters in the Monte Carlo 
constrained by MiniBooNE data.
Uncertainties on the parameters modeling the optical properties
of the oil in the MiniBooNE detector are constrained by a fit
to the calibration sample of Michel electrons.

An observation of neutrino oscillation in MiniBooNE would have corresponded to an
excess of candidate electron neutrino events over expected backgrounds.
The number of observed events in the analysis region, $475 < E_\nu < 1250$ MeV, 
was $380\pm19(stat)$. Therefore, we observed no significant excess of $22\pm19(stat)\pm35(syst)$
events. When an oscillation fit was performed, assuming $\nu_{\mu} \rightarrow \nu_{e}$ oscillation hypothesis,
we found the best fit oscillation parameters ($\Delta m^2$, $\sin^2 2 \theta$) = ($4.0$ eV$^2$, 0.001)
with 99\% probability, compared to a null oscillation hypothesis probability of 93\%.
Therefore, we found no significant excess of $\nu_e$ events for neutrino energy above 475 MeV,
suggesting that the data are inconsistent with models explaining LSND with a significant component of
$\nu_{\mu} \rightarrow \nu_{e}$ oscillations. More details may be found in Ref.~\cite{MB_osc}.

When the analysis was extended to lower energies we observed an excess of $\nu_{e}$-like events in the region 
below 475 MeV.  The difference between the data and the prediction was found to be $188\pm27(stat)\pm47(syst)$ 
events in the energy range $200 < E_\nu < 475$ MeV. 

Published explanations for the low-energy excess include 
anomaly-mediated neutrino-photon coupling \cite{hhh} and neutrino oscillations involving sterile neutrinos 
\cite{sorel,weiler,goldman,maltoni,nelson}.

\section{New Analysis Results}

This section concentrates on investigation of the low-energy electron-like events.
In the course of this investigation the data sample increased from $5.58 \times 10^{20}$ POT to $6.46 \times 10^{20}$, by including
 all available data collected in neutrino mode.
The updates in the analysis included improved measurement of NC $\pi^{0}$ events and incorporation of coherent production~\cite{MB_nu_pi0}.
Several strong interaction processes have been added to the detector model. However, only photo-nuclear interactions
had a sizable effect. These interactions can cause a photon from $\pi^{0}$ decay in the detector to be missed, leaving a single photon
than cannot be distinguished from an electron. 
The addition of photo-nuclear interactions increased the estimated background from NC $\pi^{0}$ by about 30\% in the
$200 < E_{\nu} <475$ MeV neutrino energy range.
Another improvement comes from improved measurement and rejection of the dirt events, i.e. background events at low energy caused 
by $\nu_{\mu}$ interactions in the tank wall and dirt surrounding the detector.

\begin{figure*}[htb]
\centering
\includegraphics[width=80mm,height=50mm]{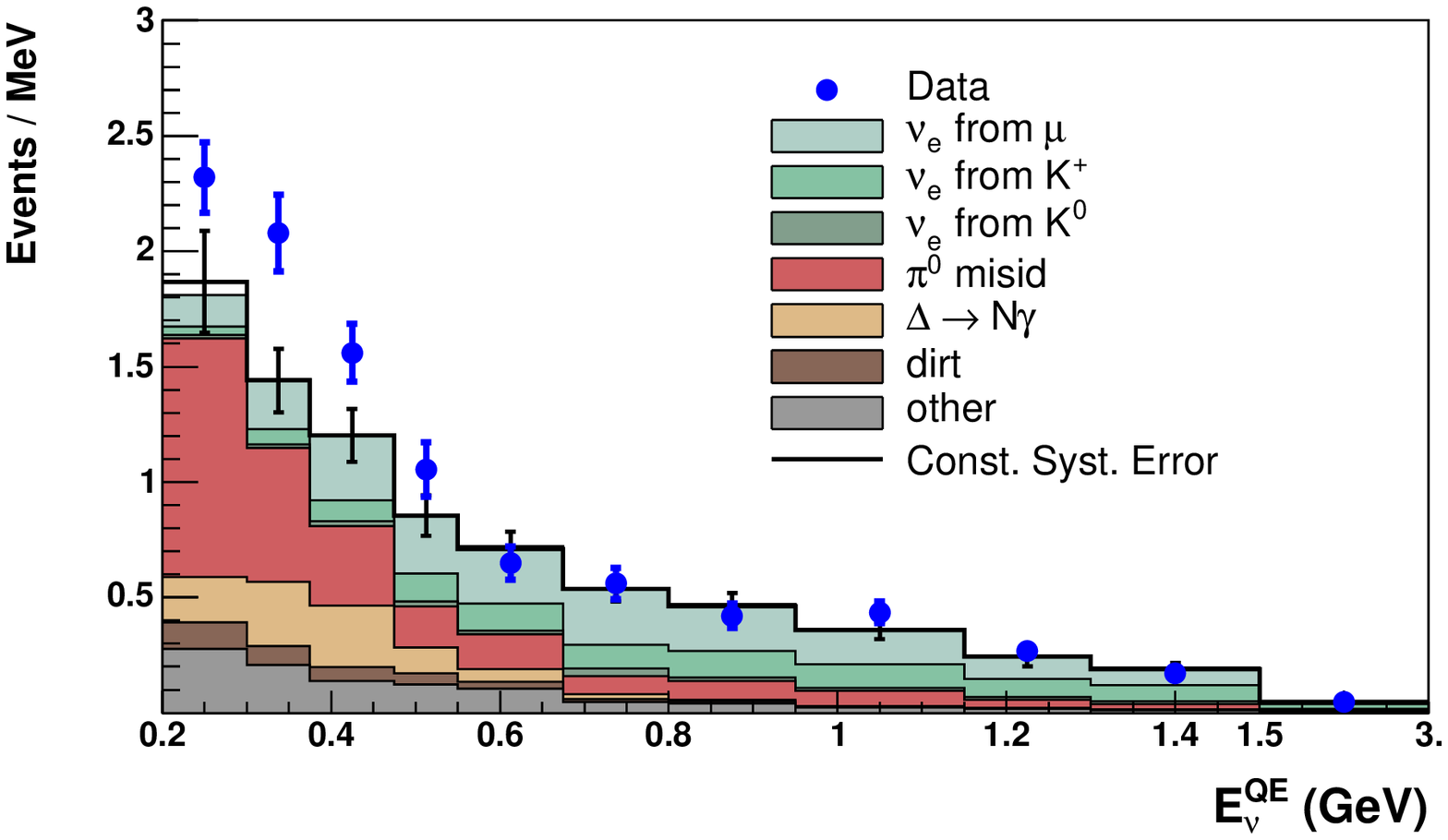}\includegraphics[width=80mm,height=50mm]{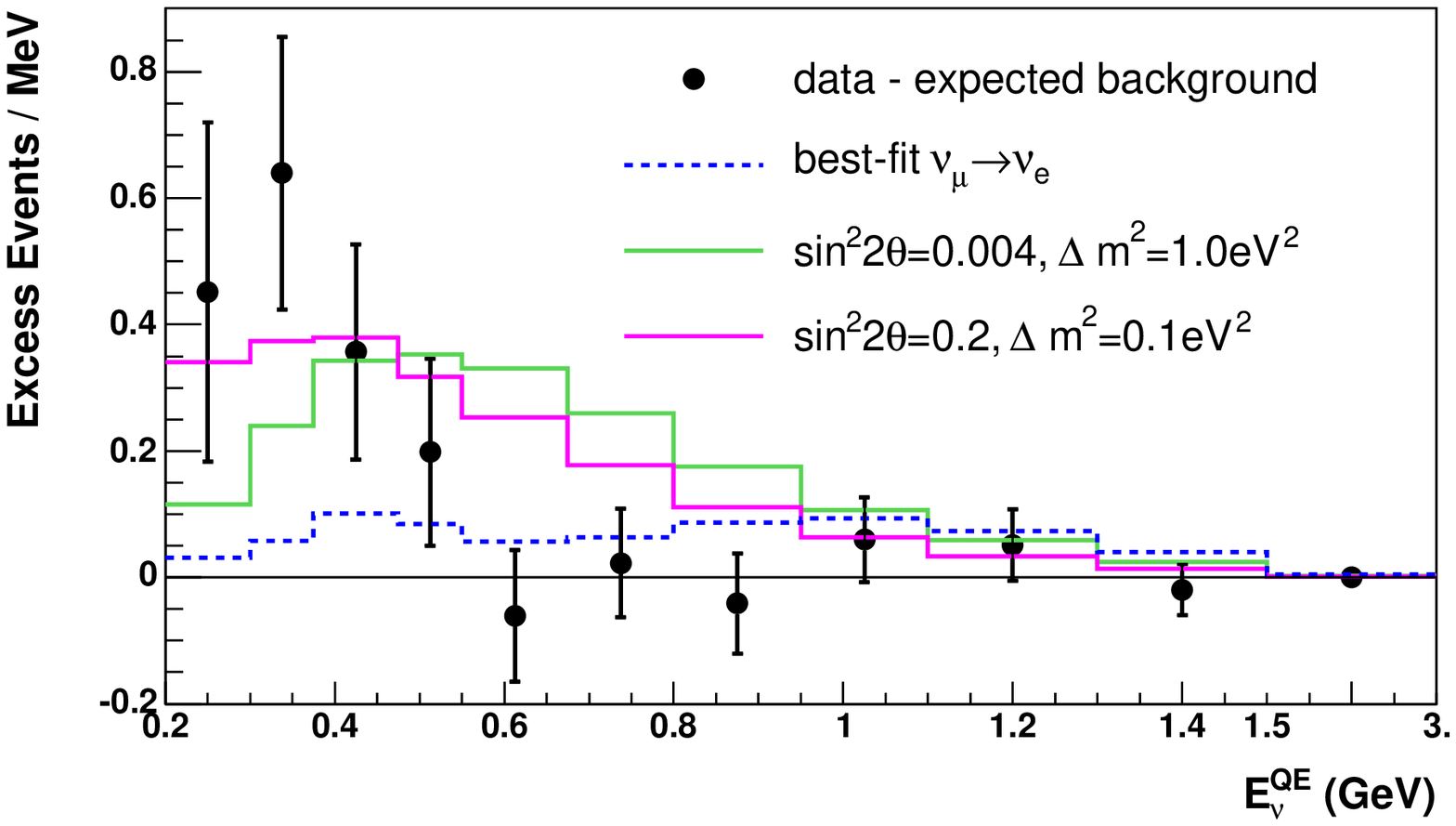}
\caption{Left: Reconstructed $E_{\nu}$ distribution of $\nu_{e}$ CCQE candidates.  The data is shown as the points with statistical error.
The background prediction is shown as the histogram with systematic errors.
Right: The difference between the data and predicted backgrounds as a function of reconstructed neutrino energy. The error bars include both
statistical and systematic components. } 
\label{fig:MBnew_figure1}
\end{figure*}

Fig.~\ref{fig:MBnew_figure1} shows reconstructed $E_{\nu}$ distribution of $\nu_{e}$ CCQE candidates (left).
The data is shown as the points with statistical error.
The background prediction is shown as the histogram with systematic errors and includes the improvements in the analysis mentioned above.
The right panel of Fig.~\ref{fig:MBnew_figure1} shows the difference between the data and predicted backgrounds as a function of reconstructed neutrino energy. The error bars include both statistical and systematic components. 

Table~\ref{table:three_BNB_nu_EnuQE_bins} shows observed data and predicted background event numbers in three $E_{\nu}$
bins. The total background is broken down into the intrinsic $\nu_{e}$ and $\nu_{\mu}$ induced components.
The $\nu_{\mu}$ induced background is further broken down into its separate components.
\begin{table}[htb]
\begin{tabular}[c]{| llll |}\hline
$E_{\nu}$[GeV]                                    & 0.2-0.3              & 0.3-0.475             & 0.475-1.25           \\\hline
Total Bkgd                                            & 186.8$\pm$26 & 228.3$\pm$24.5 & 385.9$\pm$35.7 \\
$\nu_{e}$ induced                              & 18.8                   & 61.7                       & 248.9                    \\
$\nu_{\mu}$ induced                          & 168                   & 166.6                     & 137                        \\\hline
NC $\pi^{0}$                                         & 103.5               & 77.8                       & 71.2                        \\
NC $\Delta\rightarrow N \gamma$  & 19.5                  & 47.5                       & 19.4                       \\
Dirt                                                         & 11.5                  & 12.3                       &  11.5                     \\
Other                                                     & 33.5                  & 29                           & 34.9                      \\\hline
Data                                                      & 232                   & 312                        &  408                       \\\hline
Data-MC                                              & 45.2$\pm$26  & 83.7$\pm$24.5    &  22.1$\pm$35.7  \\\hline
Significance                                        & 1.7$\sigma$    & 3.4$\sigma$         & 0.6$\sigma$       \\ \hline
\end{tabular}
\caption{Observed data and predicted background event numbers in three $E_{\nu}$
bins. The total background is broken down into the intrinsic $\nu_{e}$ and $\nu_{\mu}$ induced components.
The $\nu_{\mu}$ induced background is further broken down into its separate components.\label{table:three_BNB_nu_EnuQE_bins}}
\end{table}
Therefore, MiniBooNE observes an unexplained excess of $128.8 \pm 20.4 \pm 38.3 $ electron-like events in the energy region
$200 < E_{\nu} < 475$ MeV.
The details the analysis of the low energy electron-like events are described in~\cite{BNB_nu_lowE}. 
\begin{figure*}[htb]
\centering
\includegraphics[width=100mm,height=130mm]{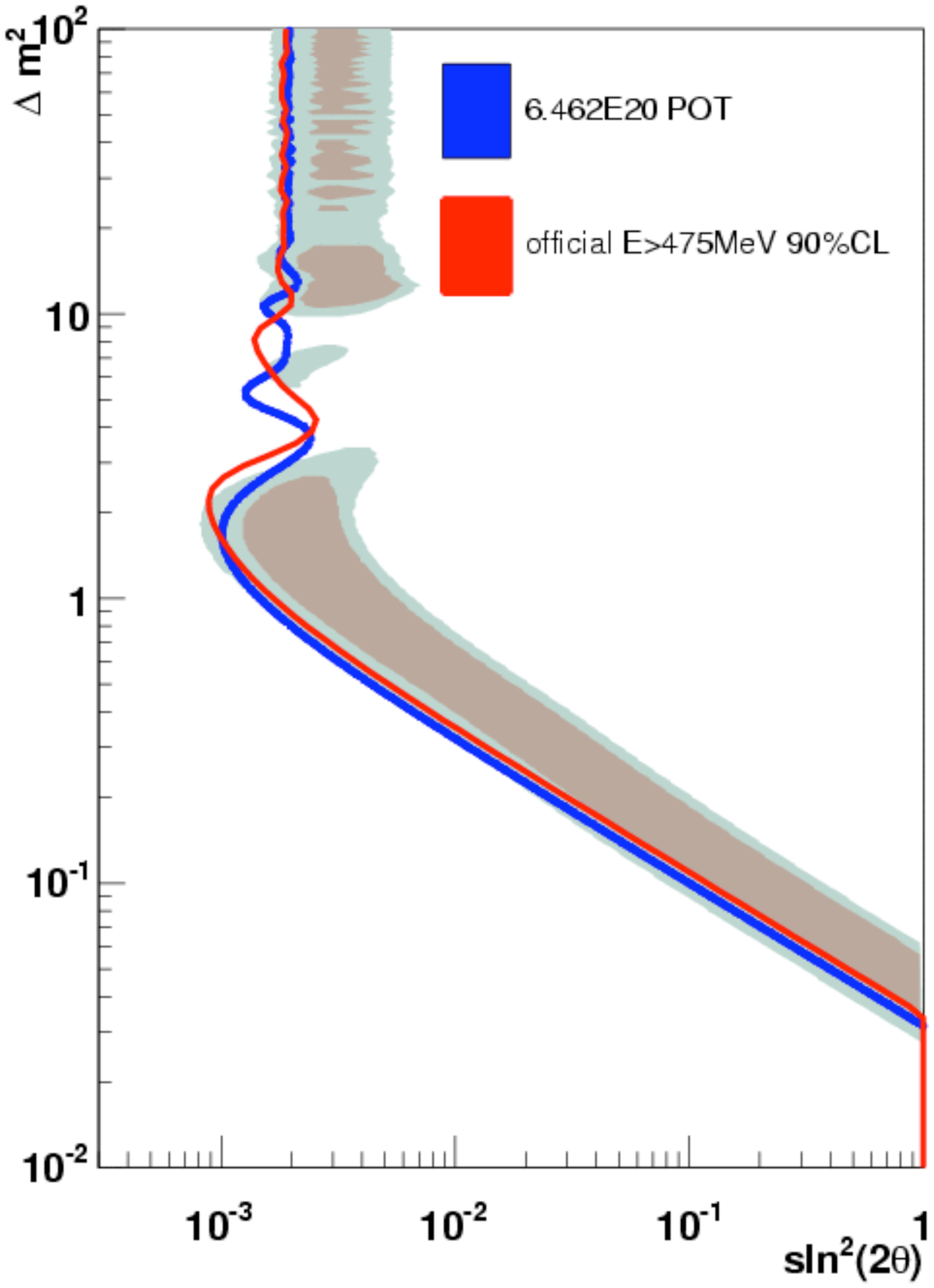}
\caption{The oscillation fit performed in the energy range 475 to 1250 MeV does not change with the updates in the analysis.
The limit to $\nu_{\mu}$ to $\nu_{e}$ oscillations obtained with 
$5.58 \times 10^{20}$ protons on target (red curve) data set, and with $6.46 \times 10^{20}$ protons on target
(blue curve) data set.
} 
\label{fig:limit}
\end{figure*}
The oscillation fit performed in the energy range 475 to 1250 MeV does not change with the updates in the analysis.
The limit to $\nu_{\mu}$ to $\nu_{e}$ oscillations is shown in Fig.~\ref{fig:limit}, with 
$5.58 \times 10^{20}$ protons on target (red curve), and with $6.46 \times 10^{20}$ protons on target
(blue curve) data set.

It is clear that more information is needed to understand the difference between the data and prediction
in the low-energy region.
Fortunately, additional handles will come from the data collected in the MiniBooNE detector with
the anti-neutrinos coming from the BNB beam line, and from the data collected in the same detector
from the NuMI neutrino beam.

\section{Events from NuMI beam line observed by MiniBooNE detector}

\begin{figure*}[htb]
\centering
\includegraphics[width=160mm,height=130mm]{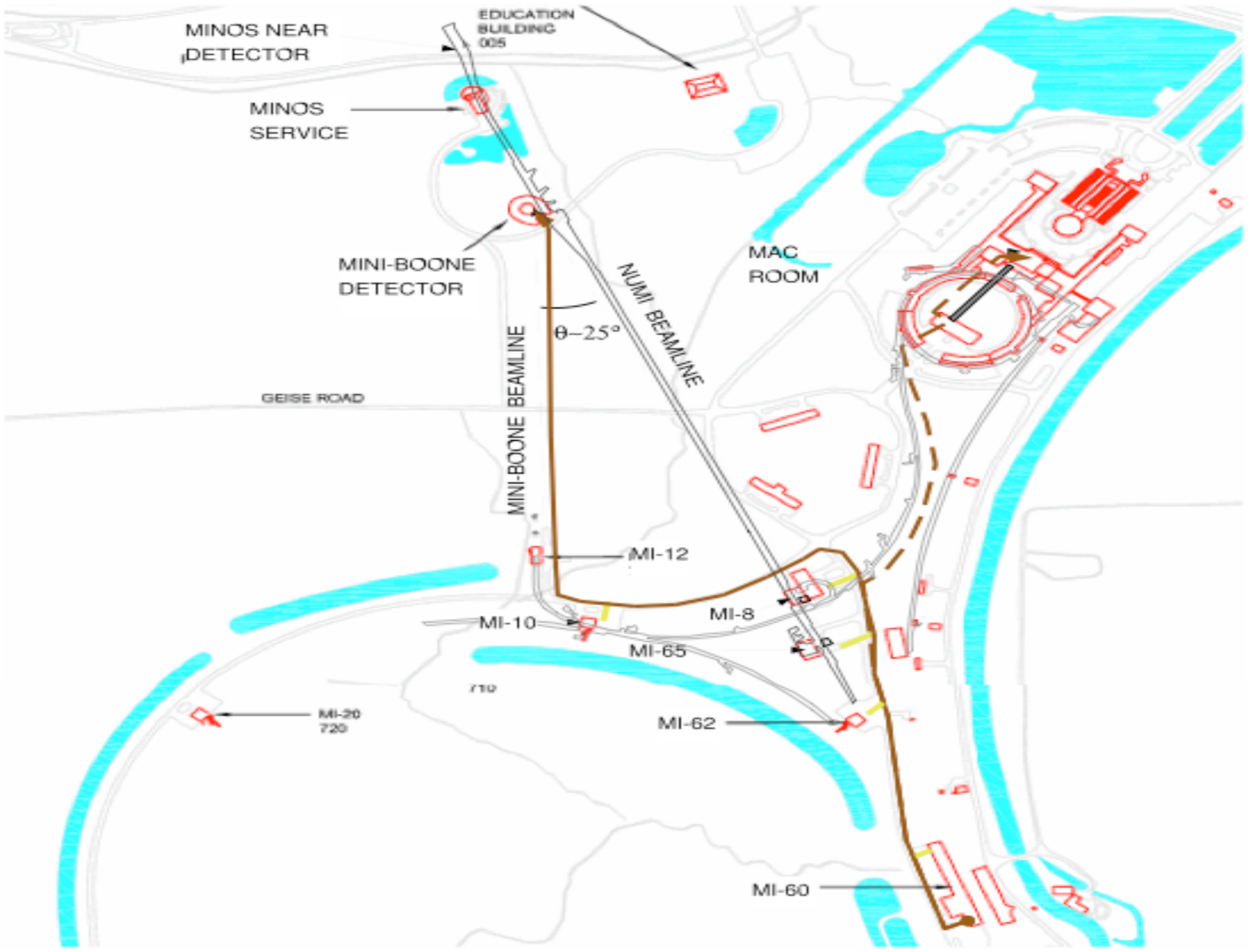}
\caption{Fermi Nation Accelerator Laboratory is currently running two beam lines producing neutrinos. The Booster Neutrino Beam is producing neutrinos
used in the MiniBooNE experiment. The NuMI Beam is emitting neutrinos intended for use in the MINOS experiment.} 
\label{fig:beamlines}
\end{figure*}

\begin{figure*}[htb]
\includegraphics[trim= 20mm 90mm 20mm 15mm, height=4.8cm, width=8.6cm]{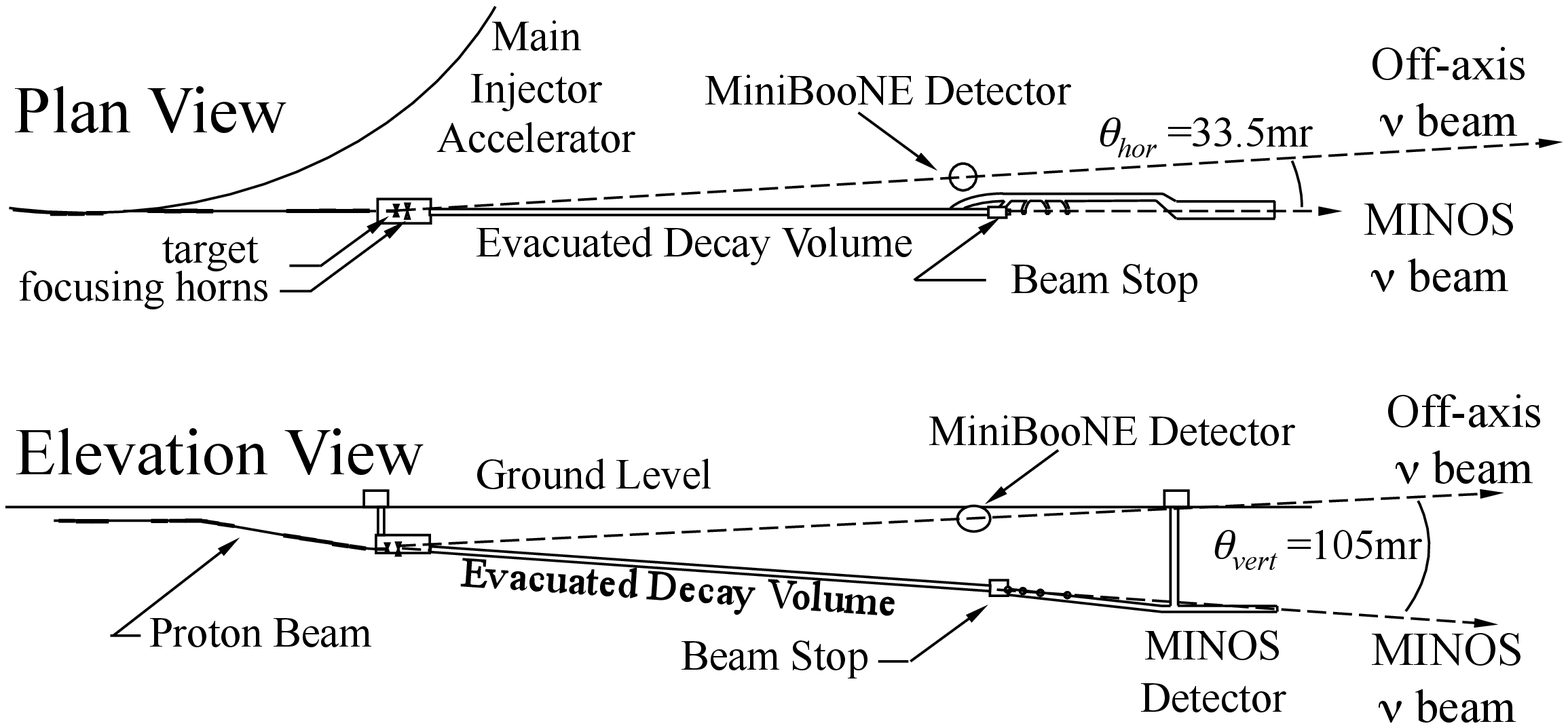}
\caption{\label{fig:NuMI_beamline} Plan and elevation views of the NuMI beam line with respect to the MiniBooNE detector. 
}
\end{figure*}

\begin{figure*}[htb]
\centering
\includegraphics[width=80mm,height=50mm]{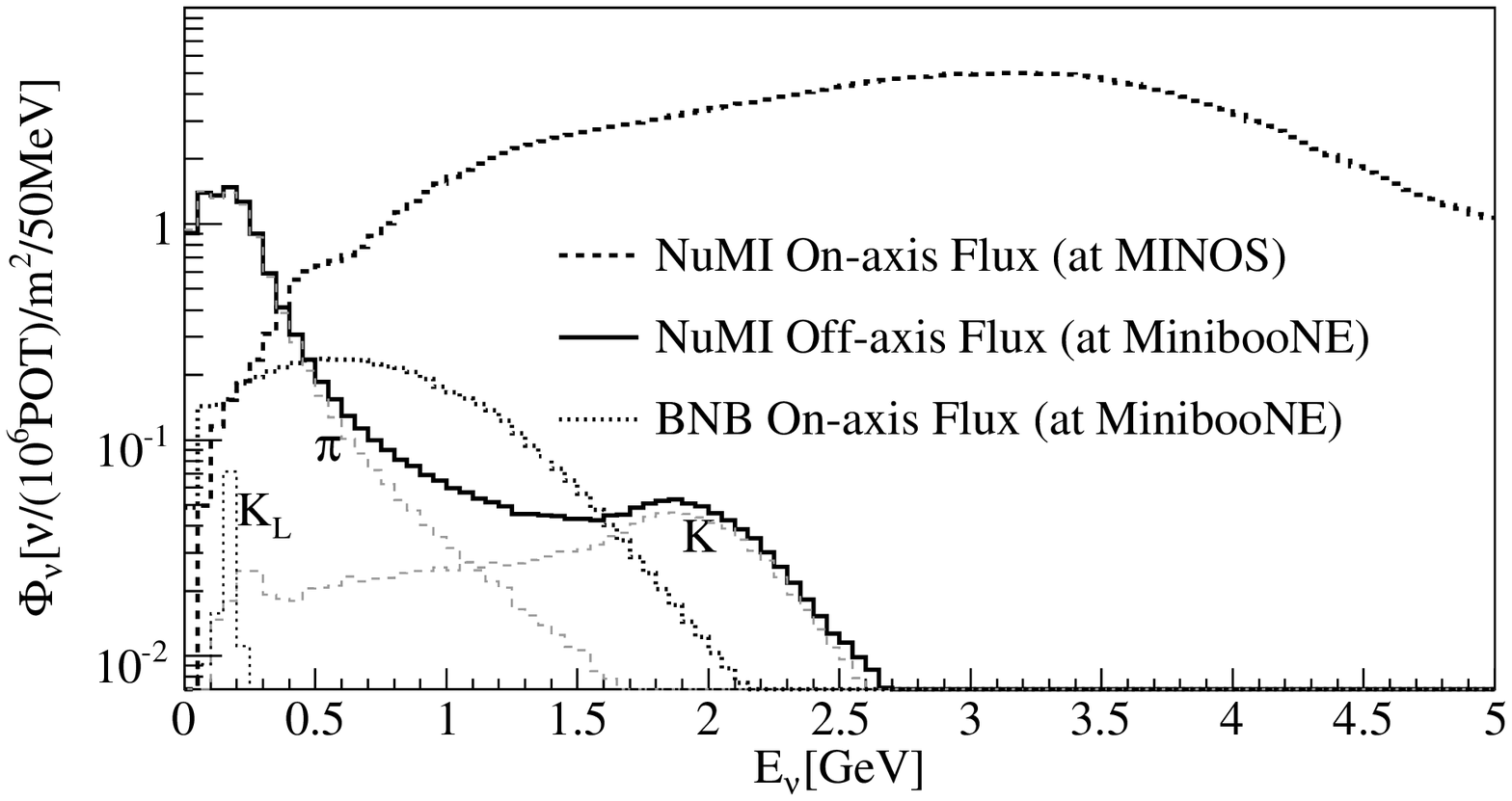}
\caption{Comparison of the predicted NuMI off-axis, NuMI on-axis, and MiniBooNE fluxes including all neutrino species.
The off-axis flux is separated into contributions from charged $\pi$ and $K$ parents. } 
\label{fig:flux_comparison}
\end{figure*}

Fermilab has two beam lines producing neutrinos: the BNB and NuMI beamilne,
as shown in Fig.~\ref{fig:beamlines}.
The MiniBooNE detector, located at an angle of 110~mrad ($6.3^{\circ}$) with respect to the NuMI beam axis,
provides a unique opportunity to perform the first measurement of neutrino interactions from an off-axis horn-focused beam. 
Future neutrino oscillation searches by the T2K~\cite{T2K} and NO$\nu$A~\cite{Nova}
experiments plan to use off-axis horn-focused beams. 
The details of the NuMI beam line with respect to the MiniBooNE detector are shown
in Fig.~\ref{fig:NuMI_beamline}.
The NuMI beam points toward the
MINOS Far Detector, located in the Soudan Laboratory in Minnesota.
Neutrinos are produced by 120~GeV protons incident on a carbon target.
Positive $\pi$ and $K$ mesons produced in the target are focused down the decay pipe using two
magnetic horns.
The NuMI neutrino flux at the MiniBooNE detector is shown in Fig.~\ref{fig:flux_comparison}.
Detailed  {\sc GEANT3}-based Monte Carlo (MC) simulations of the beam, 
including secondary particle production,
 particle focusing, and transport, are performed to calculate the flux as a
function of neutrino flavor and energy.  
The yield of pions and kaons from the NuMI
target is calculated using the {\sc FLUKA} cascade model~\cite{fluka}. 
The beam modeling includes downstream interactions in material other than the target that produce 
hadrons decaying to neutrinos.
These interactions are modeled using a {\sc GEANT3} simulation, 
configured to use either {\sc GFLUKA}~\cite{gfluka} or {\sc GCALOR}~\cite{gcalor} cascade models. 

Pions and kaons produce neutrinos with average energies of about 0.25~GeV  and 2~GeV, respectively.
The peak structure in the NuMI off-axis flux distribution is a consequence 
of the two-body decay kinematics.
The energy of $\nu_{\mu}$s from two-body decays is given by
\begin{equation}
E_{\nu}\approx{\frac{{\Big(1-{\frac{{m_{\mu}^{2}}}{{m_{\pi,K}^{2}}}
}\Big)E_{\pi,K}}}{{1+\gamma^{2} \tan^2\theta}}},
\end{equation}
where $m_{\pi,K}$ ($E_{\pi,K}$) is the mass (energy) of the $\pi$, $K$ parent, and  $m_\mu$ is the muon mass.
Therefore at a suitable off-axis angle $\theta$, the neutrino flux is confined to a relatively narrow band of energies,
which is useful in limiting backgrounds in searches for the oscillation transition $\nu_{\mu} \rightarrow \nu_e$. 

Samples of charged current quasi-elastic $\nu_{\mu}$ and $\nu_e$ interactions were analyzed. 
The high rate and simple topology of  $\nu_{\mu}$ CCQE events
provided a useful sample for understanding
the $\nu_{\mu}$ spectrum and verifying the MC prediction for 
$\nu_{e}$ production. The identification of $\nu_{\mu}$
CCQE events was based upon the detection of the primary stopping muon and the associated
decay electron as two distinct time-related clusters of PMT hits, called 'subevents':
$\nu_{\mu}+n\rightarrow\mu^{-}+p,\;\;\;\mu^{-}\rightarrow e^{-}+\nu_{\mu}+\bar{\nu}_{e}$.
\begin{figure*}[htb]
\centering
\includegraphics[width=80mm,height=50mm]{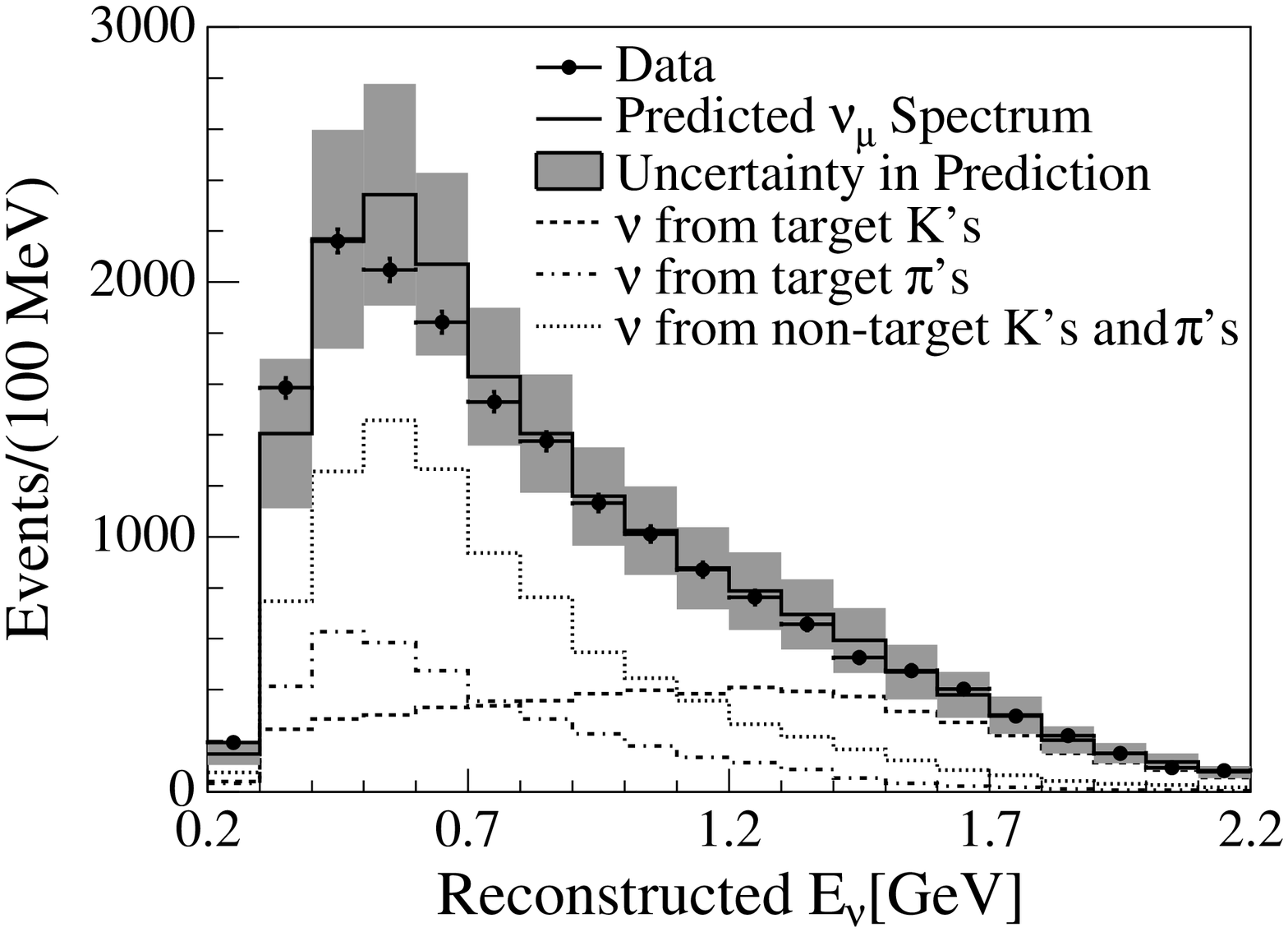}\includegraphics[width=80mm,height=50mm]{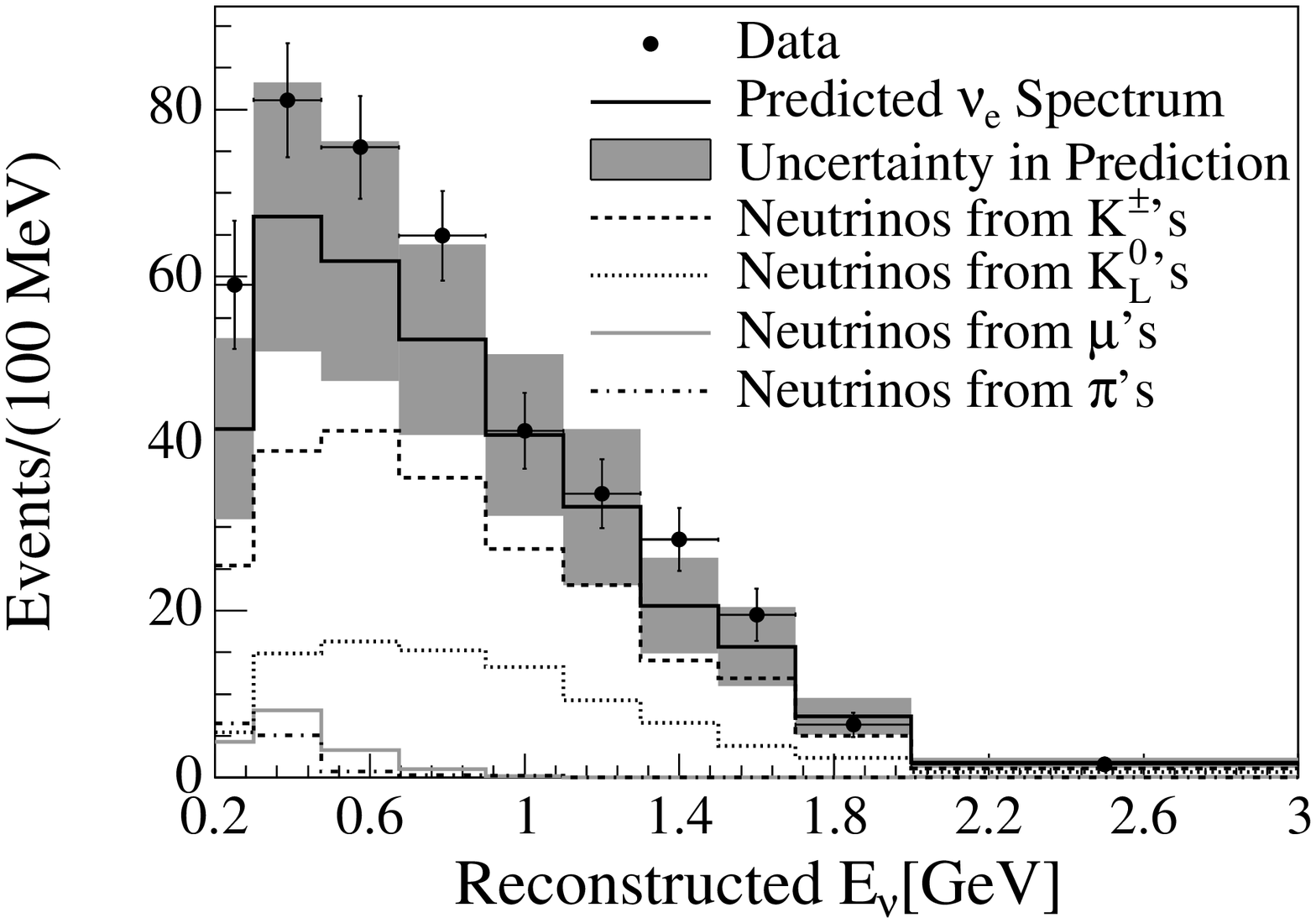}
\caption{Left: Reconstructed $E_{\nu}$ distribution of $\nu_{\mu}$ CCQE events.
The band indicates the total systematic uncertainty associated with the MC prediction.
Also shown are the contributions from pion (52\%) and kaon (48\%) parents of neutrinos. 
Right: Reconstructed $E_{\nu}$ distribution of $\nu_{e}$ CCQE candidates. 
The prediction is separated into contributions from neutrino parents.
The band indicates the total systematic uncertainty associated with the MC prediction. 
Kaon parents contribute 93\% of the events in this sample.} 
\label{fig:numu_and_nue}
\end{figure*}
The reconstructed neutrino energy, $E_{\nu}$, distribution of selected $\nu_{\mu}$
CCQE events is shown in Fig.~\ref{fig:numu_and_nue}, along with the MC prediction.
A total of 17659 data events passed this $\nu_{\mu}$ CCQE selection criteria,
compared to the prediction of 18247$\pm$3189; the uncertainty includes systematic errors
associated with the neutrino flux, neutrino cross-sections,
and detector modeling.
The flux uncertainties included particle production in the NuMI target, modeling of the downstream
interactions, and kaons stopped in the NuMI beam dump.
The $\pi/K$ yields were tuned to match the MINOS Near Detector data~\cite{adamson} in several of the NuMI beam configurations.
 Such tuning has a negligible effect on the off-axis beam at MiniBooNE.
However, the difference between untuned and tuned 
$\pi/K$ yields is treated as 
an additional  systematic effect.
The cross-section uncertainties and the uncertainties in the parameters describing the
optical properties of the MiniBooNE detector are quantified in the way described in the analysis
of the BNB events. The agreement between data and the prediction of the neutrino flux from $\pi/K$ parents 
indicates that the NuMI beam modeling provides a good description of the observed off-axis $\nu_\mu$ 
flux in MiniBooNE.

\begin{figure*}[htb]
\centering
\includegraphics[width=80mm,height=50mm]{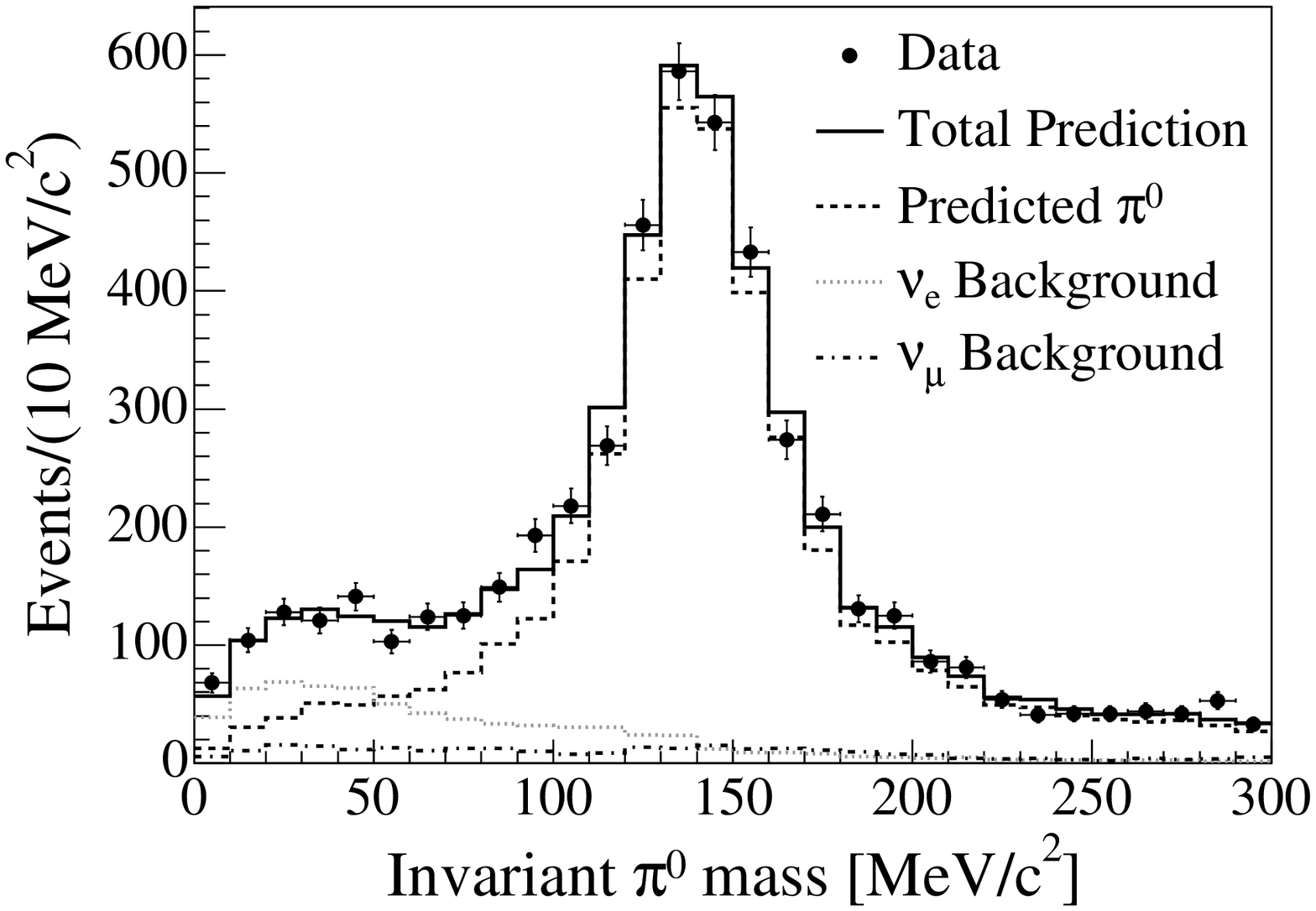}\includegraphics[width=80mm,height=49mm]{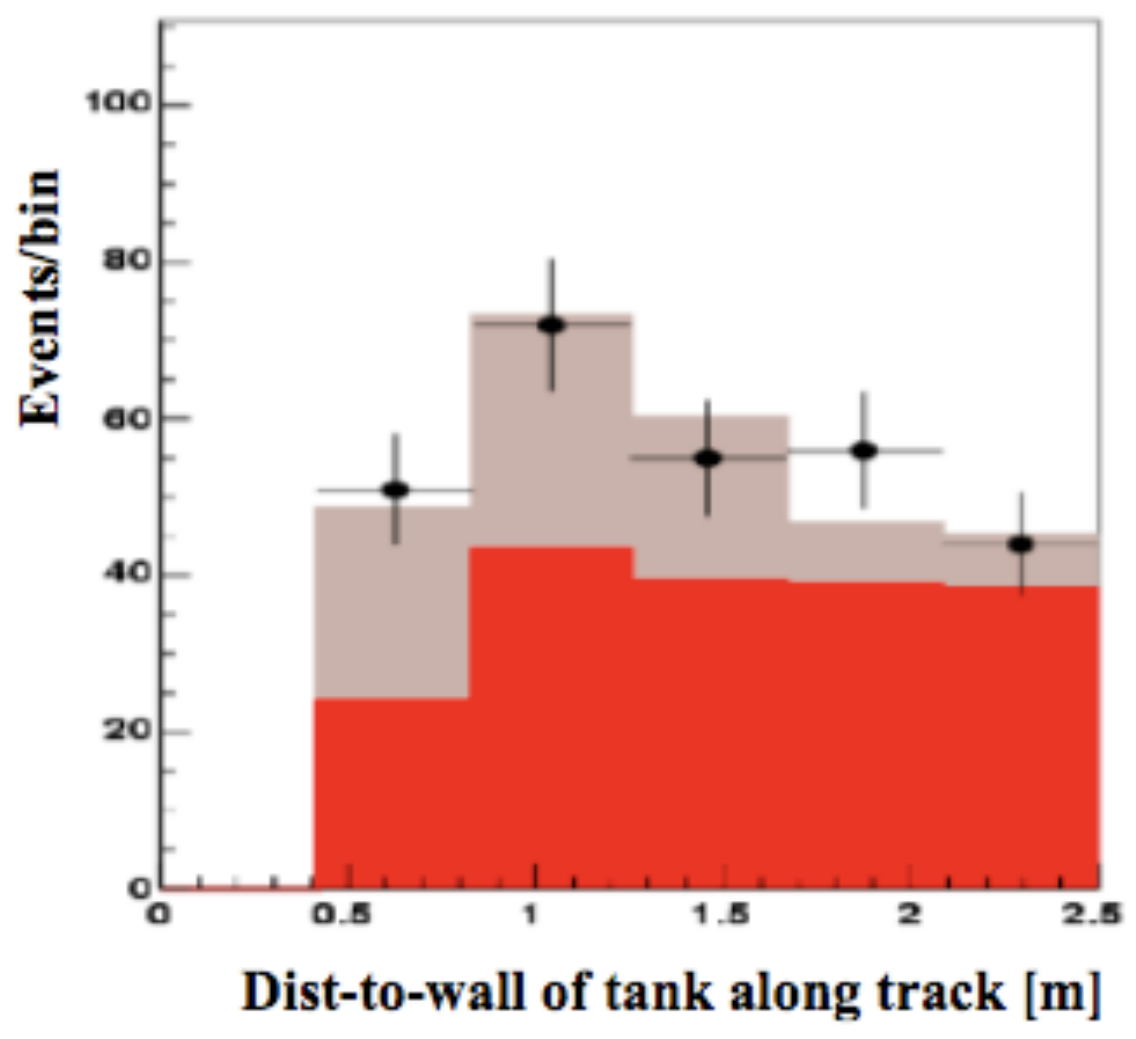}
\caption{Left: Mass distribution of NC $\pi^{0}$ candidates for data (points) and MC (solid histogram).
The dashed histogram is the subset of predicted events with at least one true $\pi^{0}$. 
Predicted non-$\pi^{0}$ backgrounds are either from $\nu_{\mu}$ and $\bar{\nu}_{\mu}$ (dash-dotted line) or $\nu_{e}$ and $\bar{\nu}_{e}$ (dotted line) interactions. 
Kaon parents contribute 84\% of the events in this sample.
Right: The distribution of he distances from the detector wall measured along the track the particle created.
The events are separated in the components originating outside the detector (shown in brown), and the component originating within the detector (shown in red).
} 
\label{fig:pi0_and_dirt}
\end{figure*}

The $\nu_e$ CCQE events consist of a single subevent of PMT hits 
($\nu_{e}+n\rightarrow e^{-}+p$).
The majority of the remaining background
is NC $\pi^{0}$ and the dirt events with only a single reconstructed electromagnetic track
that mimics a $\nu_{e}$ CCQE event. 
To test our NC  $\pi^{0}$ prediction, a clean sample of NC $\pi^{0}$ events is reconstructed, as shown in Fig.~\ref{fig:pi0_and_dirt} (left).
This sample demonstrates good agreement between data and MC. 
Fig.~\ref{fig:pi0_and_dirt}(right) shows the sample of the dirt events that originate from outside of the MiniBooNE detector.
The variable shown is the distance from the detector wall measured along the track the particle created.
The events are separated in the components originating outside the detector (shown in brown), and the component originating within the detector (shown in red).
This distribution shows good agreement between data and MC and therefore shows that the
backgrounds at low energy that result from neutrino interactions in the 
tank wall and dirt surrounding the detector are understood.
A total of 780 data events pass all of the $\nu_{e}$ CCQE selection criteria.
The MC prediction is 660$\pm$129 
with a  $\nu_{e}$ CCQE efficiency of 32\% and purity of 70\%. 
The corresponding energy distribution is shown in Fig.~\ref{fig:numu_and_nue} (right side).

To facilitate further comparison, the low energy and high energy regions 
are divided at 0.9 GeV. 
The number of data and Monte Carlo events in these two energy regions is given in
Table~\ref{table:two_NuMI_EnuQE_bins_nue_cut200}. 
\begin{table}[ptb]
\begin{tabular}[c]{| l|ll |}\hline
$E_{\nu}$[GeV]           & 0.2-0.9              & 0.9-3.0                    \\ \hline
Total Bkgd                    & 401$\pm$78    & 259$\pm$56         \\
$\nu_{e}$ induced     & 311                    & 231                         \\
$\nu_{\mu}$ induced & 90                      & 28\\\hline
NC $\pi^{0}$               & 28                       & 25\\
NC $\Delta\rightarrow N \gamma$& 14& 1\\
Dirt                               & 36                       &    1\\
Other                           & 12                       & 1                                  \\ \hline
Data                            & 496$\pm$22     & 284$\pm$17             \\ \hline
Data-MC                    & 95$\pm$81        & 25$\pm$58\\ \hline
\end{tabular}
\caption{Observed data and predicted background event numbers in two $E_{\nu}$
bins. The total background is broken down into the intrinsic $\nu_{e}$ (and
$\bar{\nu}_{e}$) and $\nu_{\mu}$ (and $\bar{\nu}_{\mu}$) induced components.
The $\nu_{\mu}$ (and $\bar{\nu}_{\mu}$) induced background is further broken
down into its separate components.\label{table:two_NuMI_EnuQE_bins_nue_cut200}}
\end{table}
There is an indication of an excess in the data at above 1 $\sigma$ level
in the region of reconstructed neutrino energy $E_{\nu}<0.9$ GeV.
The details are given in Ref.~\cite{first_NuMI}. 

Therefore, both samples of charged current quasi-elastic $\nu_{\mu}$ and $\nu_e$ interactions are found to be in agreement with expectation. 
This directly verifies the expected pion and kaon contributions to the neutrino flux and validates the modeling of the NuMI off-axis beam. 
In particular, the $\nu_{\mu}$ CCQE
sample demonstrates an excellent understanding of the details of both the pion
and kaon contributions to the neutrino beam. The $\nu_{e}$ CCQE sample also agrees 
with the prediction, but with some indication of a slight excess for neutrino energies below $0.9$ GeV.
In addition to demonstrating the off-axis beam technique, the measurement verifies
the predicted fluxes from $\pi/K$ parents in the NuMI beam, and probes the off-axis intrinsic $\nu_e$ contamination,
required for future $\nu_\mu \rightarrow \nu_e$ appearance searches.

After the demonstration of the off-axis concept, useful in limiting backgrounds in searches for the oscillation transition $\nu_{\mu} \rightarrow \nu_e$, the analysis
is currently directed toward examing the low-energy region and searching for oscillation.
In this way it will complement the analysis done with MiniBooNE using neutrino and anti-neutrino BNB data, but with different systematics.
It is worth noting that the NuMI $\nu_{e}$ CCQE sample has a very different composition
when compared to the BNB neutrino $\nu_{e}$ CCQE sample. The BNB $\nu_{e}$ sample
originates mostly from decays of pions produced in the target, and contains large fraction
of $\nu_{\mu}$ mis-identified events. On other hand, the NuMI $\nu_{e}$ CCQE sample is produced
mostly from decay of kaons, and contains a dominant fraction of intrinsic $\nu_{e}$ events.
The analysis will be done by forming a correlation between the large statistics $\nu_{\mu}$ CCQE sample 
and $\nu_{e}$ CCQE, and by tuning the prediction to the data simultaneously. This is a method equivalent to forming a ratio between 
near and far detectors in two-detector experiments where the near detector detects $\nu_{\mu}$ CCQE events, 
while the far detector samples $\nu_{e}$ CCQE events. The result is that the prediction is being tuned to 
the data, and common systematics cancel; this might reveal something profound about the nature
of the $\nu_e$ sample. 

\section{Conclusion}
The MiniBooNE experiment detected an unexplained excess of $128.8 \pm 20.4 \pm 38.3$ electron-like events in reconstructed 
neutrino energy range from 200 to 475 MeV. The excess might originate either from an unknown background component, or could be explained
with a new physics process.
The NuMI data sample currently has a large systematic errors associated with  $\nu_{e}$ events,
but shows an indication of a similar excess.
Incoming analysis of the NuMI data with constrained systematic errors, and anti-neutrino data sample collected with
the Booster beam will help distinguish various possibilities.

\begin{acknowledgments}
I would like to acknowledge the support of Fermilab, the Department of Energy, and
the National Science Foundation. 
\end{acknowledgments}

\end{document}